\documentclass[fleqn,usenatbib,useAMS]{mnras}

\usepackage{graphicx}
\usepackage{amsmath}
\usepackage{amssymb}    
\usepackage{multicol}      
\usepackage{bm}
\usepackage{float}
\usepackage{chngcntr}
\usepackage{soul}
\usepackage{pdflscape}    
\usepackage{caption}
\usepackage{subcaption}
\usepackage{enumitem}
\usepackage{array}
\usepackage[toc,page]{appendix}
\usepackage{xcolor}
\usepackage{threeparttable}
\usepackage{xspace}

\newcommand{\kms}{\mbox{$\mathrm{km\,s^{-1}}$}}
\newcommand{\Ha}{\mbox{${\mathrm H\alpha}$}}

\newcommand{\Ion}[2]{#1{\,\scriptsize #2}}

\newcommand{\Lines}[3]{\Ion{$[$#1}{#2}$]$\,$\lambda\lambda$\,#3}



\title[MRS-N S147]{Decomposing the Kinematics of SNR G180.0-01.7 with Medium-Resolution Spectra}
\author[Ren et al.]{Juan-Juan Ren$^{1}$\thanks{Contact e-mail: \href{mailto:jjren@nao.cas.cn}{jjren@nao.cas.cn}}, 
Wei Zhang$^{1,2}$,
Chao-Jian Wu$^{1,2}$, 
Chao Liu$^{1}$, 
Jian-Jun Chen$^{1}$, 
Hong Wu$^{1,2}$\\
$^{1}$ National Astronomical Observatories, Chinese Academy of Sciences, Beijing 100101, China\\
$^{2}$ CAS Key Laboratory of Optical Astronomy, National Astronomical Observatories, Chinese Academy of Sciences, Beijing 100101, China
}

\date{Accepted 2026 September 22. Received 2026 September 2; in original form 2026 June 12}

\begin{document}
\maketitle

\begin{abstract}

The exceptionally large angular diameter and intrinsic optical faintness of the Galactic supernova remnant (SNR) G180.0-01.7 (S147) have historically hindered a comprehensive spectroscopic census of its filamentary structure. We present the first dedicated medium-resolution (R\,$\sim$\,7500) spectroscopic study of S147 using the LAMOST Medium-Resolution Spectroscopic survey of Galactic Nebulae (MRS-N). By implementing multi-Gaussian decomposition on the \Ha, [\Ion{N}{II}], and [\Ion{S}{II}] emission complexes and applying combined spectroscopic and spatial criteria, we disentangle the shock-ionized filaments of the SNR from the pervasive foreground and background Diffuse Ionized Gas (DIG). Our analysis reveals a distinct bimodal radial velocity distribution for the SNR: a low-velocity population tracing the decelerated forward shock at a dense cavity wall, and a high-velocity population (with absolute radial velocities of $\sim$\,70\,$\pm$\,30\,\kms) corresponding to ejecta within a low-density cavity. The co-existence of these populations provides kinematic evidence for the ``supernova-in-a-cavity'' scenario. Furthermore, we detect enhanced [\Ion{S}{II}]/\Ha\ ratios in the DIG along the eastern and northern boundaries of the remnant, signaling shock compression of the ambient interstellar medium and highlighting remnant's multifaceted feedback. 

\end{abstract}

\section{Introduction}

Supernova remnants (SNRs) serve as the primary laboratories for investigating the chemical enrichment, energy injection, and cosmic-ray acceleration that govern the evolution of the interstellar medium (ISM) and galactic ecology \citep{Long2017hsn..book.2005L, Bamba2022hxga.book...77B}. Despite their fundamental importance, the observed population of Galactic SNRs remains profoundly incomplete. Current catalogs list only $\sim$\,310 confirmed remnants \citep{Green2025JApA...46...14G}, revealing a stark deficit relative to the thousands predicted by stellar population synthesis models \citep{Ranasinghe2022ApJ...940...63R}. This ``missing SNR'' problem primarily stems from selection effects. While distant young remnants are often missed due to limited angular resolution, highly evolved remnants with large angular diameters tend to fade below the surface brightness limits of traditional surveys, rendering them indistinguishable from the pervasive diffuse ionized gas (DIG) that dominates optical emission of the Galactic plane.

SNR G180.0-01.7 (S147, also known as the Spaghetti Nebula) is a prime example of such a challenging evolved remnant. Located toward the Galactic anticenter, it is one of the largest and most well-known shell-type remnants, exhibiting a nearly spherical morphology with an angular diameter of $\sim$\,200\arcmin \citep{Minkowski1958RvMP...30.1048M, van1960ZA.....51...15V}. Multi-wavelength observations have painted a complex portrait: radio maps reveal a circular shell structure \citep{Dickel1969ApJ...155...67D, Reich2003A&A...408..961R, Xiao2008A&A...482..783X}, while X-ray observations with SRG$/$eROSITA detect soft thermal emission coincident with the optical shell but also reveal morphological asymmetries \citep{Michailidis2024A&A...689A.277M,Khabibullin2024A&A...689A.278K}. Furthermore, S147 harbors the radio pulsar PSR J0538+2817, which exhibits a rare 3D spin-velocity alignment, thereby providing a unique anchor for studying the supernova explosion mechanism \citep{Yao2021NatAs...5..788Y}. And S147 is also of interest as it is the best candidate for containing an unbound surviving companion, OB star HD 37424 \citep[$\sim$\,13\,M\sun;][]{Dincel2015MNRAS.448.3196D}, which is the pre-supernova binary companion to the progenitor of the pulsar \citep{Cruz-Cruz2026ApJ...999..162C}.

A fundamental obstacle to understanding S147 has been the substantial uncertainty in its distance, which propagates into errors in its physical size, age, and energy budget. Early estimates varied wildly from 0.6 to 1.9 kpc based on different methods \citep{Clark1976MNRAS.174..267C, Fesen1985ApJ...292...29F, Guseinov2003A&AT...22..273G, Kramer2003ApJ...593L..31K, Chatterjee2009ApJ...698..250C, Dincel2015MNRAS.448.3196D, Chen2017MNRAS.472.3924C}. This impasse was recently resolved by \citet{Kochanek2024ApJ...968...94K}, who utilized spectra of luminous blue stars to measure absorption-line distances to the remnant. They firmly established a revised distance of 1.37 kpc (with a 90 per cent confidence interval of 1.30\,--\,1.47 kpc), a value consistent with the median parallax distance to the associated pulsar PSR J0538+2817. This precise distance provides the essential foundation for a re-evaluation of the remnant's physical scale and evolutionary status.

Furthermore, a fundamental tension persists regarding the remnant's age and evolutionary stage. The characteristic age of PSR J0538+2817 ($\sim$\,0.62\,Myr) is an order of magnitude larger than a kinematic age of $\sim$\,37.6\,kyr inferred from the pulsar's proper motion and its offset from the putative explosion center \citep{Khabibullin2024A&A...689A.278K}. This stands in stark contrast to the $\sim$\,150\,kyr Sedov expansion age estimated for the SNR shell \citep{Lozinskaya1976SvA....20...19L,Khabibullin2024A&A...689A.278K}. This disparity among these age estimates constitutes the ``age dilemma'' of S147 \citep{Reich2003A&A...408..961R}. A natural resolution is provided by the ``supernova-in-a-cavity'' scenario, wherein the explosion occurs within a low-density cavity excavated by the progenitor's stellar wind. In this framework, the forward shock propagates freely through the tenuous interior before being significantly decelerated upon encountering the dense swept-up shell at the cavity wall, thereby reconciling a large physical radius with a substantially younger kinematic age \citep{Chevalier1989ApJ...344..332C,Dwarkadas2005ApJ...630..892D,Dwarkadas2007ApJ...667..226D}. 

This scenario is grounded in a well-established theoretical framework: the canonical wind-blown bubble (WBB) model \citep{Castor1975ApJ...200L.107C,Weaver1977ApJ...218..377W} predicts a dense outer shell whose radial expansion follows  $R$\,$\propto$\,$t^{3/5}$ during the energy-conserving phase, and subsequent studies by \citet{Chevalier1989ApJ...344..332C} and \citet{Dwarkadas2005ApJ...630..892D,Dwarkadas2007ApJ...667..226D} indicate that an SNR expanding into such a pre-structured cavity can deviate substantially from the classical Sedov-Taylor evolution, with the Sedov phase potentially shortened or bypassed completely. Specifically for S147, \citet{Khabibullin2024A&A...689A.278K} recently argued that the remnant's X-ray properties are broadly consistent with the ``supernova-in-a-cavity" scenario, which naturally resolved the age discrepancy. In their picture, a massive progenitor \citep[21.5\,--\,41.4 M$_{\sun}$;][]{Cruz-Cruz2026ApJ...999..162C} excavated a low-density wind-blown bubble, within which the subsequent explosion occurred.

However, a complete picture of S147 must also account for its distinctive and asymmetric morphology. The remnant exhibits prominent ``ear-like'' protrusions in the east and west \citep{Grichener2017MNRAS.468.1226G, Ren2018RAA....18..111R}. The origin of these features is debated, being attributed either to jet-driven shaping during core-collapse explosion \citep{Soker2023RAA....23k5017S} or to the interaction of the forward shock with an inhomogeneous circumstellar medium. Distinguishing between these scenarios requires precise kinematical constraints, which the ``supernova-in-a-cavity'' framework is well-suited to provide. Thus, testing this model also directly addresses the enigma of the remnant's asymmetric morphology. Furthermore, the detection of an unbound massive companion HD~37424 suggests that the progenitor of S147 was part of a binary system \citep{Dincel2015MNRAS.448.3196D,Cruz-Cruz2026ApJ...999..162C}. This raises the possibility that the circumstellar environment of the SN progenitor was shaped not only by its own wind but also by the wind of its companion, potentially leaving an imprint on the observed morphology of the SNR \citep{Soker2023RAA....23k5017S,Ercolino2025A&A...696A.103E,Ercolino2026A&A...706A.169E}.

Resolving both the age dilemma and the origin of the asymmetric morphology in S147 therefore necessitates a comprehensive set of precise, spatially resolved kinematic measurements of its shocked gas. Historically, such data have been scarce. Early spectroscopic studies were limited to a few bright filaments \citep{Kirshner1979ApJ...229..147K, Fesen1985ApJ...292...29F}, while even recent large-scale low-resolution survey lack the spectral resolution to decouple the blended emission lines from the SNR filaments and the pervasive DIG. This has left the expansion velocity and true kinematic structure of S147 uncertain, and the cavity scenario largely untested at the kinematic level. Moreover, multi-wavelength evidence suggests that the feedback of S147 extends beyond the optical shell. Recent $\gamma$-ray observations reveal escaping cosmic rays interacting with adjacent molecular clouds \citep{Yang2026ApJ..1009...10Y}, highlighting the profound impact of the remnant on its surrounding medium. 

To overcome these limitations, we present the first dedicated medium-resolution (R\,$\sim$\,7500) spectroscopic study of S147 using data from the LAMOST Medium-Resolution Spectroscopic survey of Galactic Nebulae \citep[MRS-N;][]{Wu2021RAA....21...96W}. The enhanced resolution allows for the first robust decomposition of SNR shock emission from the DIG, enabling precise measurements of radial velocities and line ratios across the vast extent of the remnant. In this paper, we leverage the unique capabilities of the LAMOST MRS-N to obtain the robust, spatially-resolved kinematics of S147. We therefore aim to: (1) test the predictions of the ``supernova-in-a-cavity'' scenario by searching for its kinematic signatures, thereby providing independent evidence to resolve the age dilemma; and (2) characterize the shock properties and feedback of the remnant on its local environment. The structure of this paper is as follows. Section 2 describes the observations and data reduction. Section 3 presents the analysis of the emission-line fitting, the combined spectroscopic and spatial separation of SNR and DIG, and the kinematic mapping. Section 4 discusses the jet-driven nature of the eastern ear, the ``supernova-in-a-cavity'' dynamical scenario, and the remnant's feedback on the ambient ISM. Finally, we summarize our conclusions in Section 5.

\section{Observations and data reduction}
\label{sec:observations}

The Large sky Area Multi-Object fiber Spectroscopic Telescope (LAMOST, also known as Guo Shou Jing Telescope) is a 4-meter effective aperture quasi-meridian reflecting Schmidt telescope, which is equipped with 4000 fibers on its 5$^\circ$ diameter field of view focal plane \citep{Wang1996ApOpt..35.5155W, Su2004ChJAA...4....1S, Cui2012RAA....12.1197C, Zhao2012RAA....12..723Z}. As a dedicated spectroscopic survey telescope, LAMOST has completed its first five years low spectral resolution (R\,$\sim$\,1800) regular survey (i.e. the first stage survey program) on 2017. Since September 2017, the second stage survey program of LAMOST (i.e. LAMOST-II) was initiated. Besides the low-resolution spectroscopic observations, LAMOST-II began the medium-resolution spectroscopic survey \citep[MRS;][]{Liu2020arXiv200507210L} by using the bright/gray nights (from the 7th to the 22nd on the lunar calendar) every lunar month. LAMOST-II MRS provides spectra with a medium-resolution of R\,$\sim$\,7500 \citep{Liu2020arXiv200507210L}, covering of 4950\,--\,5350\,\AA\ (blue channel) and 6300\,--\,6800\,\AA\ (red channel). Depending on the scientific aims, the MRS survey includes several different sub-surveys, one of them is the Galactic Nebulae survey \citep[MRS-N;][]{Wu2021RAA....21...96W}. 

The main scientific goals of MRS-N survey focus on \Ion{H}{II} regions, supernovae remnants (SNRs), planetary nebulae (PNe) and Herbig-Haro objects \citep{Wu2021RAA....21...96W}. The relatively higher resolution and dedicated wavelength coverage (covering the nebulae emission lines \Ha, \Lines{N}{II}{6548,\,6584}, and \Lines{S}{II}{6717,\,6731}) make MRS-N an ideal spectroscopic survey to sample the optical emission lines of Galactic nebulae \citep{Wu2021RAA....21...96W}. MRS-N survey mainly focuses on the emission nebulae on the Galactic plane (40$^\circ$\,$<$\,l\,$<$\,215$^\circ$, $-$5$^\circ$\,$<$\,b\,$<$\,5$^\circ$), totally about 1700 deg$^2$. 

Until February 2024, 11 plates observed by MRS-N have overlaps with S147. The observational data of S147 are reduced using MRS-N's dedicated data reduction pipeline \citep{Wu2022RAA....22g5015W}, which includes removing cosmic rays, merging single exposure spectra, fitting skylight emission lines and wavelength re-calibration \citep{Ren2021RAA....21...51R}, as well as \Ha\ sky line subtraction \citep{Zhang2021RAA....21..280Z}. After rejecting the stellar spectra and invalid bad spectra, the remaining emission-line spectra are used to do fitting and parameter measurements.

\begin{figure*}
\begin{center}
\includegraphics[scale=0.60]{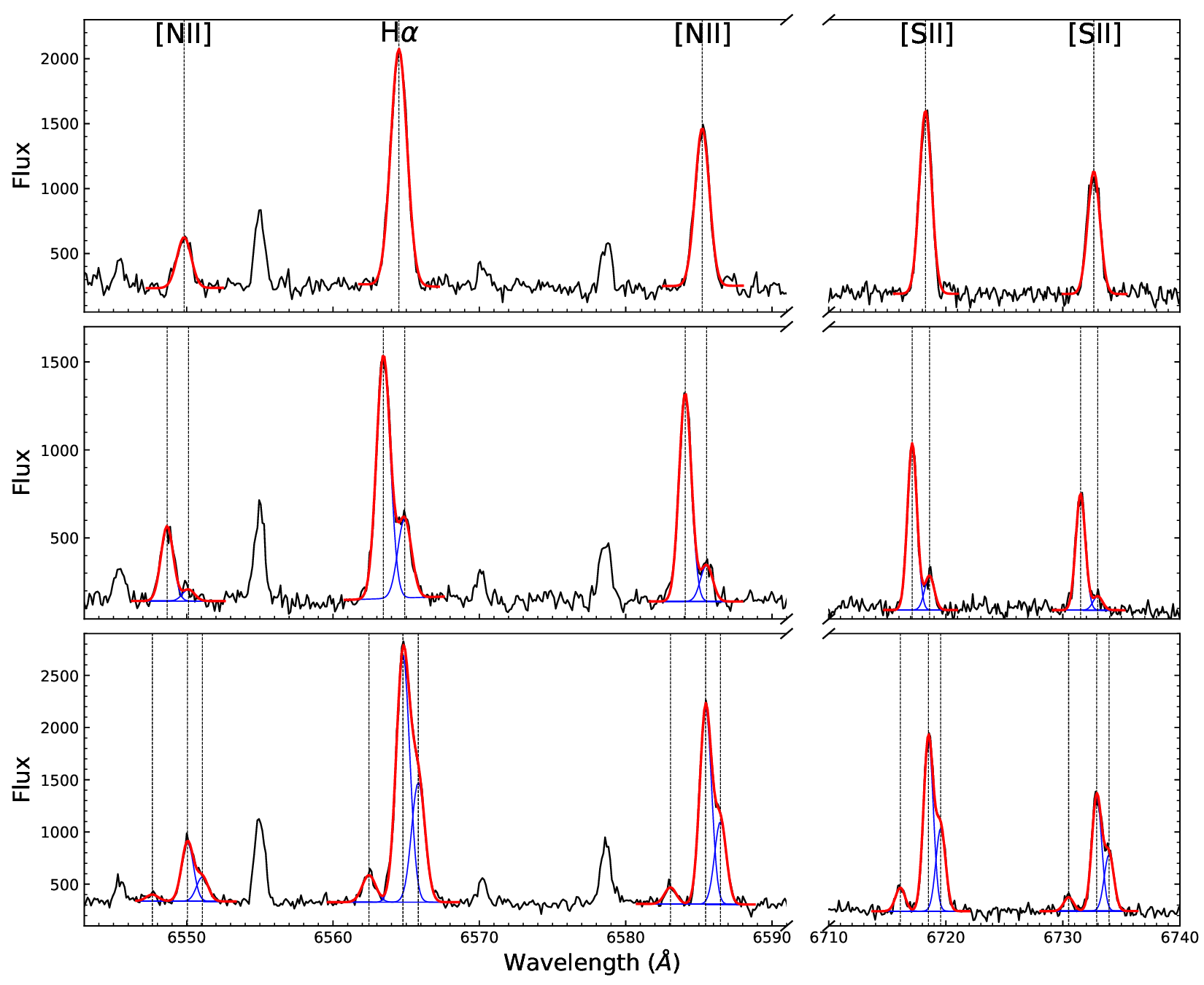} 
\caption{The fitting examples for single-component (upper panels), double-components (middle panels) and three-components (bottom panels) emission-line spectra. The red curves show the Gaussian fitting results, with vertical black lines the fitted line centers. The blue curves in the middle and bottom panels show the decomposed multi-Gaussian components. Note that the geocoronal \Ha\ emission is subtracted, the other emission lines in the spectra are skylines which are not subtracted as they have no overlap with \Ha, \Lines{N}{II}{6548,\,6584}, and \Lines{S}{II}{6717,\,6731} we used.}
\label{fig:example_fitting}
\end{center}
\end{figure*}

\section{Analysis and results}

\subsection{Emission-line fitting and Parameter measurements}

Due to the large angular size of S147 and the complexity of its filamentary structure, the optical emission is often contaminated by foreground/background DIG and suffers from line blending. Visual inspection of the spectra revealed that many emission lines exhibit multi-peaked profiles. These complex profiles arise from the superposition of emission from the SNR's shock-ionized filaments, and the surrounding DIG. The medium-resolution of LAMOST MRS-N is critical for resolving these blended components and enabling precise kinematic measurements. To de-convolve these blended components, we employed a multi-Gaussian fitting algorithm using the  {\sc lmfit} package \citep{newville_matthew_2014_11813}. The number of Gaussian components for each emission-line complex was determined dynamically by visual inspection of the line profile complexity, ranging from 1 to 3 components per spectrum. Each Gaussian component $i$ was parameterized by amplitude ($A_i$, flux), center ($\mu_i$, the central wavelength, used to derive the RV), and width ($\sigma_i$).

Following the successful convergence of the fitting routine, we extracted the physical parameters for each Gaussian component: the RV, line ratios Flux(\Lines{N}{II}{6548,\,6584})/Flux(\Ha) (hereafter [\Ion{N}{II}]/\Ha) and Flux(\Lines{S}{II}{6717,\,6731})/Flux(\Ha) (hereafter [\Ion{S}{II}]/\Ha). For details of the fitted parameters for each Gaussian component, see table 1 in \citet{Wu2022RAA....22g5015W}. The fitted RVs are referenced to the heliocentric frame. To convert them to the Local Standard of Rest (LSR) frame, we adopted the solar peculiar motion relative to the LSR as $(U_{\odot}, V_{\odot}, W_{\odot}) = (10.6, 10.7, 7.6)$ km s$^{-1}$, which are the recommended values for the A5 universal rotation curve model based on the latest VLBI maser parallax measurements \citep{Reid2019ApJ...885..131R}. All RVs discussed hereafter are reported in the LSR frame (i.e., V$_\text{LSR}$).

\subsection{Separation of SNR and Diffuse Ionized Gas}

The optical emission from S147 is contaminated by foreground/background DIG, which dominates the ionized ISM in the Galactic plane. Accurate separation of SNR shock-ionized gas from DIG is critical to characterizing the remnant's kinematics and physical conditions, as their emission-line properties may overlap at low surface brightness.

\subsubsection{Physical Characteristics of DIG}

DIG (also termed the warm ionized medium) constitutes about 90 percent of the Milky Way's ionized gas \citep{Reynolds1971PhDT.........1R} and is characterized by low electron density, high electron temperature, and enhanced ratios of collisionally-excited to recombination emission lines compared to classical \Ion{H}{II} regions \citep{Madsen2006ApJ...652..401M}. Unlike \Ion{H}{II} regions powered by OB stars, DIG ionization is attributed to non-stellar sources: leakage of photons from star-forming regions \citep{Haffner2009RvMP...81..969H}, shocks from supernova winds \citep{Collins2001ApJ...551...57C}, cosmic rays \citep{Vandenbroucke2018MNRAS.476.4032V}, or hot low-mass evolved stars \citep{Binette1994A&A...292...13B}. Its diffuse morphology and elevated [\Ion{S}{II}]/\Ha\ ratio \citep[0.3\,--\,1.0;][]{Wen2025AJ....169...95W} often mimic low-surface-brightness SNRs \citep{Blair1997ApJS..108..261B,Long2022ApJ...929..144L,Li2024A&A...690A.161L}, creating contamination in unresolved spectra.

\subsubsection{Diagnostic Criteria for Separation}

Although DIG exhibits a relatively strong [\Ion{S}{II}]/\Ha\ ratio and can be confused with SNR emission at low surface brightness, the shock-ionized filaments of an SNR are expected to possess distinct kinematic signatures and a confined spatial morphology \citep{Galarza1999AJ....118.2775G,Points2019ApJ...887...66P,Li2024A&A...690A.161L}. We therefore employ a three-tiered diagnostic approach combining line-ratio diagnostics, kinematic signatures, and spatial morphology to disentangle the SNR and DIG components, leveraging the medium resolution of LAMOST MRS-N to resolve blended profiles.

\begin{figure*}
\begin{center}
\includegraphics[scale=0.55]{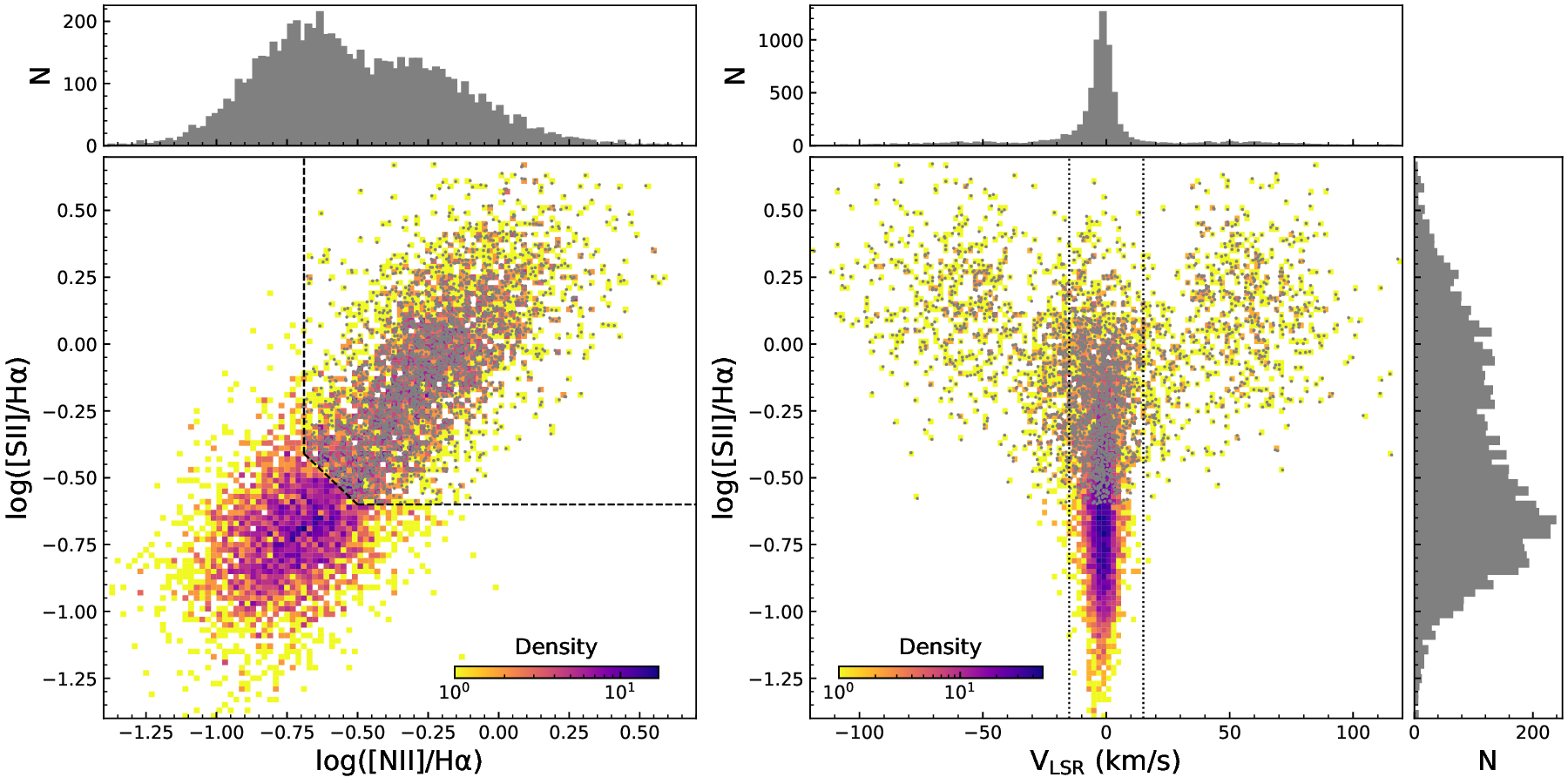} 
\caption{The log([\Ion{S}{II}]/\Ha) verse log([\Ion{N}{II}]/\Ha) (left panel) and log([\Ion{S}{II}]/\Ha) verse V$_\text{LSR}$ (right panel) plots for all the fitted emission-line parameters within 3.5 degrees of the S147 geometric center, overlaid with the number density contours. The upper and right sub-panels show the corresponding histogram distributions. The dashed lines in the left panel present the cutting criteria we used to differentiate DIG from SNR. The dotted lines in the right panel show that the DIG are mainly focused on the low V$_\text{LSR}$ region (-15 to 15 km s$^{-1})$). Components satisfying these spectroscopic criteria are further refined using a spatial boundary (see Section 3.2.2 and Figure\,\ref{fig:fig_S147}) to define the final SNR sample. The gray dots show the final separated SNR sample.}
\label{fig:fig_differentiate_snr-dig}
\end{center}
\end{figure*}

\begin{figure}
\begin{center}
\includegraphics[scale=0.45]{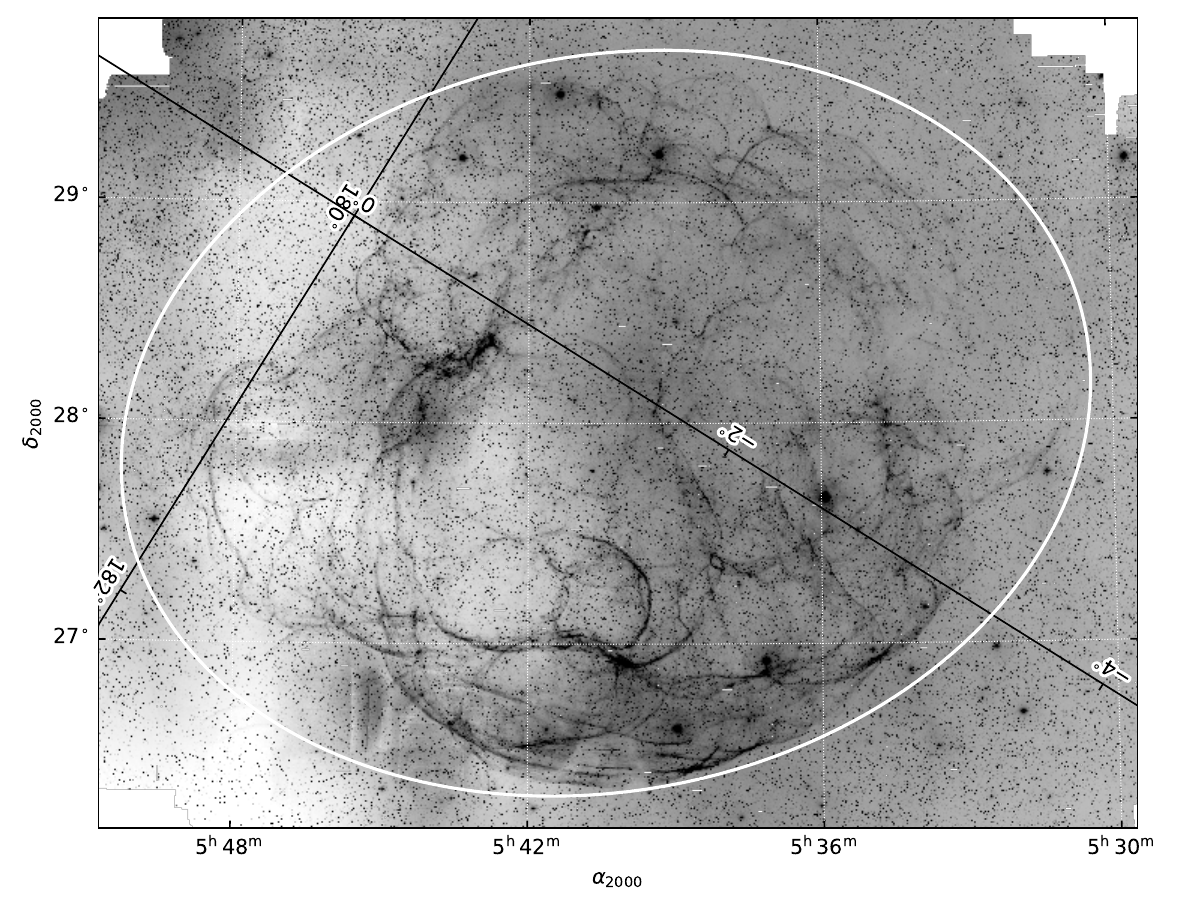} 
\caption{Spatial boundary for SNR sample selection. The \Ha\ emission map of S147 from the IGAPS survey is shown in grayscale. The white ellipse with parameters (RA$_0$, Dec$_0$, $a$, $b$, PA) = (85.1$^\circ$, $+$28.0$^\circ$, 2.5$^\circ$, 1.67$^\circ$, 172.0$^\circ$) defines the morphological boundary adopted to refine the SNR sample. }
\label{fig:fig_S147}
\end{center}
\end{figure}

\begin{figure*}
\begin{center}
\includegraphics[scale=0.50]{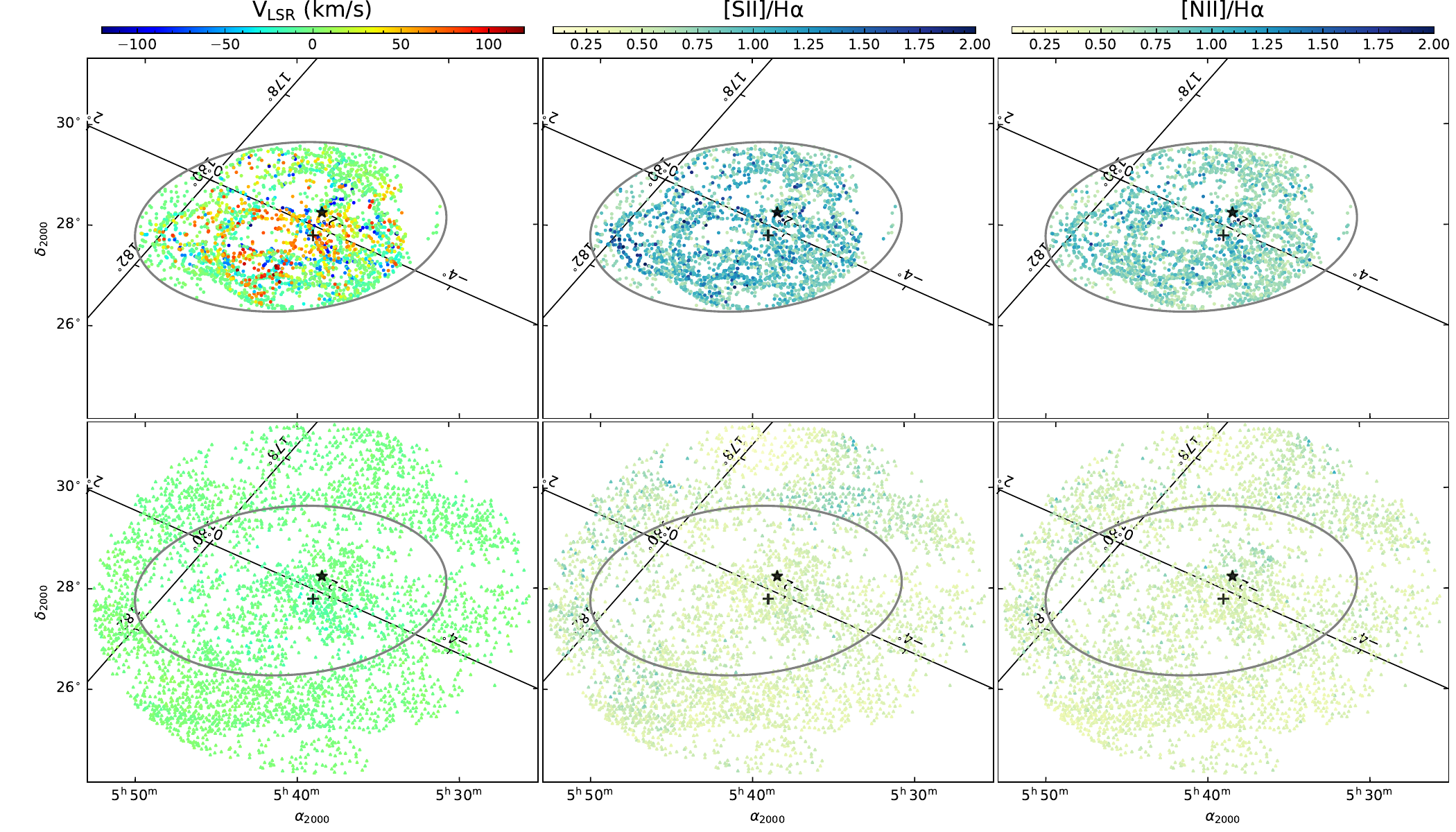} 
\caption{Upper panels: Spatial distribution of  V$_\text{LSR}$, [\Ion{S}{II}]/\Ha, and [\Ion{N}{II}]/\Ha\ for emission-line components classified as the SNR after applying both the spectroscopic criteria and the spatial boundary defined in Figure\,\ref{fig:fig_S147}. Lower panels: Same, but for components classified as the DIG (i.e., those that do not satisfy the spectroscopic shock criteria or lie outside the spatial boundary). The black asterisk and cross mark the location of PSR J0538+2817 and the geometric center of S147 \citep{Dincel2015MNRAS.448.3196D} respectively.}
\label{fig:fig_spatial_distribution_snr-dig}
\end{center}
\end{figure*}

Figure\,\ref{fig:fig_differentiate_snr-dig} (left panel) presents the diagnostic log([\Ion{S}{II}]/\Ha) versus log([\Ion{N}{II}]/\Ha) diagram for all fitted emission-line components within  3.5$^\circ$ of S147's geometric center. This angular radius is adopted because it encompasses the main \Ha\ filamentary structure visible in the narrow-band image (see Figure\,\ref{fig:fig_S147} and \ref{fig:fig_spatial_distribution_snr-dig}). Beyond this boundary, no significant filamentary structure and shock emission is detected, thereby ensuring that the gas physically associated with the SNR is included in the analysis (see also Figure\,2 in \citet{Khabibullin2024A&A...689A.278K} and Figure\,1 in \citet{Shishkin2025ApJ...992..190S}). Two distinct clusters emerge in Figure\,\ref{fig:fig_differentiate_snr-dig}: the DIG-dominated regime characterized by lower [\Ion{S}{II}]/\Ha\ and [\Ion{N}{II}]/\Ha\ ratios, and an SNR-dominated regime occupying the high line-ratio domain. To empirically separate these populations, we adopt the following spectroscopic criteria (marked by dashed lines in Figure\,\ref{fig:fig_differentiate_snr-dig}, left panel): 

\begin{align*}
\log\left(\frac{[\text{S\,{II}}]}{\Ha}\right) &> -1.0 \times \log\left(\frac{[\text{N\,{II}}]}{\Ha}\right) - 1.1, \\
\log\left(\frac{[\text{S\,{II}}]}{\Ha}\right) &> -0.60\ (\text{i.e.,\ }\frac{[\text{S\,{II}}]}{\Ha} > 0.55), \\
\log\left(\frac{[\text{N\,{II}}]}{\Ha}\right) &> -0.69\ (\text{i.e.,\ }\frac{[\text{N\,{II}}]}{\Ha} > 0.50).
\end{align*}

The right panel of Figure\,\ref{fig:fig_differentiate_snr-dig} further validates this separation kinematically. Components meeting the above line-ratio criteria (putative SNR) exhibit a broad V$_\text{LSR}$ distribution, while those that do not (putative DIG) are kinematically quiescent, clustering tightly around V$_\text{LSR}$\,$\sim$\,0\,\kms. 
 
As there maybe residual ambiguity persists in the [\Ion{S}{II}]/\Ha\ range 0.4\,--\,0.6, we introduce a morphological spatial criterion. We define the global envelope of S147 using the \Ha\ narrow-band image from the INT/WFC Galactic Plane Surveys (IGAPS) survey \citep{Greimel2021A&A...655A..49G}. An elliptical boundary with parameters (RA$_0$, Dec$_0$, $a$, $b$, PA) = (85.1$^\circ$, $+$28.0$^\circ$, 2.5$^\circ$, 1.67$^\circ$, 172.0$^\circ$) is shown in Figure\,\ref{fig:fig_S147}, which encompass the optical filamentary structures of the remnant. Only emission-line components that satisfy the spectroscopic criteria above and have fiber positions located within this elliptical boundary are retained in the final SNR sample for subsequent kinematic analysis. 

This combined spectroscopic and spatial decomposition effectively isolates the shock-ionized filaments of S147. Figure\,\ref{fig:fig_spatial_distribution_snr-dig} illustrates the results: the retained SNR components trace the remnant's filamentary shell structure, while the excluded DIG components appear diffuse and uniformly distributed across the field. This refined sample provides a robust foundation for analyzing the remnant's intrinsic expansion kinematics and shock properties.

\subsection{Kinematic Structure of the SNR}

\begin{figure*}
\begin{center}
\includegraphics[scale=0.60]{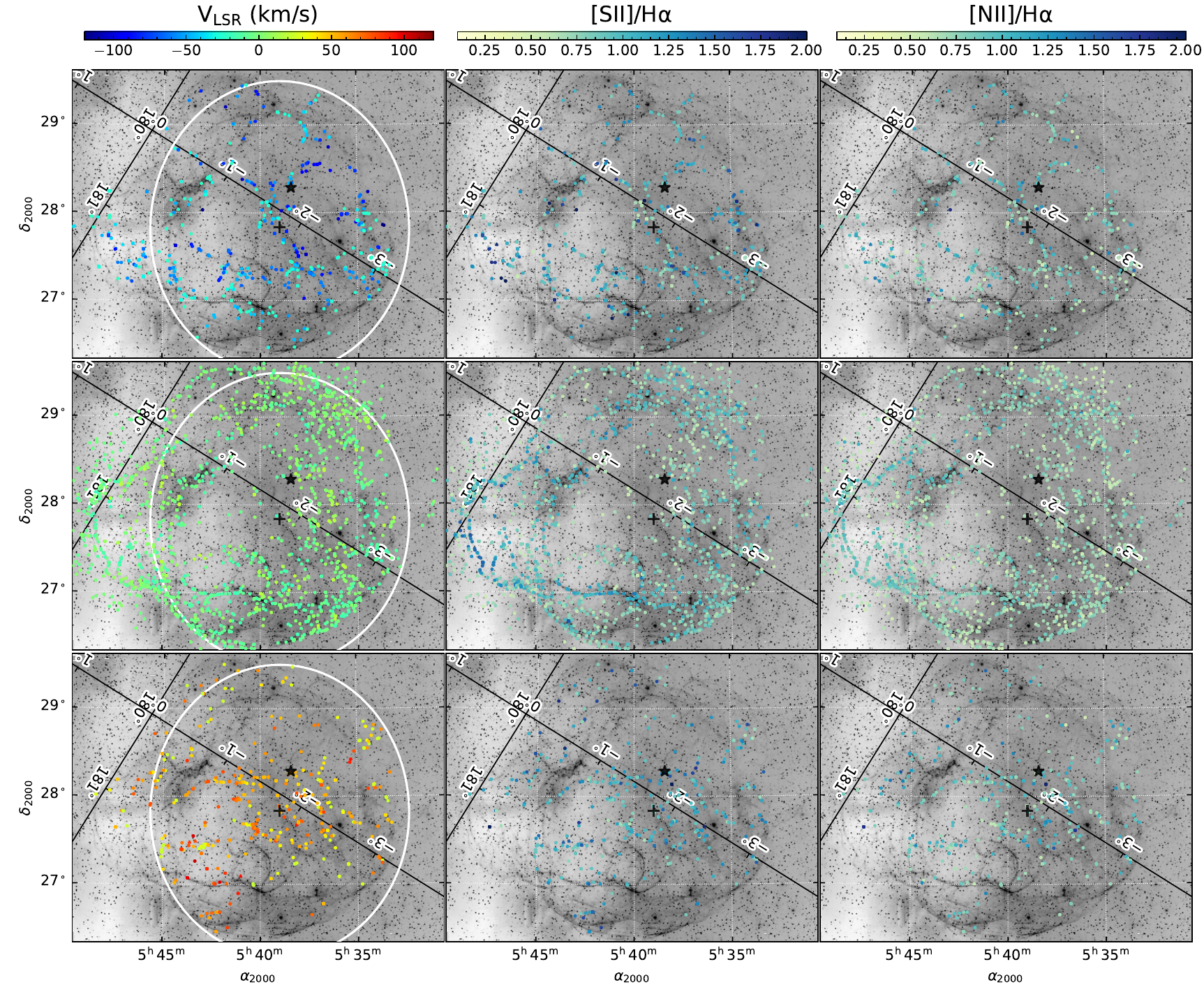} 
\caption{The spatial distribution of V$_\text{LSR}$, [\Ion{S}{II}]/\Ha, and [\Ion{N}{II}]/\Ha\ for blue-shifted (V$_\text{LSR}$\,$<$\,-20\,\kms), middle ($|$V$_\text{LSR}|$\,$\leq$\,20\,\kms), and red-shifted (V$_\text{LSR}$\,$>$\,20\,\kms) component of SNR  (from top to bottom)  respectively. Overlaid is the \Ha\ mosaic of S147 from the IGAPS survey. The black asterisk and cross mark the location of PSR J0538+2817 and the geometric center of S147 \citep{Dincel2015MNRAS.448.3196D} respectively. The white circle in the first column mark the 100 arcmin radius used to define the main shell extent.}
\label{fig:fig_spatial_distribution_snr-blue-middle-red}
\end{center}
\end{figure*}

Figure\,\ref{fig:fig_spatial_distribution_snr-blue-middle-red} shows spatial distribution of SNR components in three velocity regimes: blue-shifted (V$_\text{LSR}$\,$<$\,-20 \kms, top), intermediate-velocity ($|$V$_\text{LSR}$$|$\,$<$\,20\,\kms, middle), and red-shifted (V$_\text{LSR}$\,$>$\,20\,\kms, bottom). The multi-component decomposition described in Section 3.1 allows for a detailed reconstruction of the kinematic architecture of S147, free from the confounding effects of foreground and background DIG. The V$_\text{LSR}$ distribution of the spectroscopically separated SNR components reveals a complex, non-uniform expansion pattern. 

\subsubsection{Bimodal Velocity Distribution}

Figure\,\ref{fig:fig_RV-position} illustrates the kinematic structure of SNR G180.0-01.7. The left panel presents the velocity-position diagram, plotting the radial distance against the V$_\text{LSR}$. In this panel, the black solid line indicates a 100 \kms\ spherically expanding shell with a radius of 100 arcmin. The data points are color-coded to distinguish different regions: gray dots represent emission-lines within 100\,arcmin, black crosses mark those from the eastern ear region ($R$\,$>$\,100\,arcmin, 27$^\circ$ $<$\,DEC\,$<$\,28.5$^\circ$), while blue crosses represent the northern boundary ($R$\,$>$\,100\,arcmin, DEC\,$>$\,28.5$^\circ$). The right panel displays the corresponding histogram distribution of V$_\text{LSR}$. Analysis of the V$_\text{LSR}$ histograms (right panel of Figure\,\ref{fig:fig_RV-position}) reveals a pronounced bimodal distribution, a feature inconsistent with the predictions of a standard, homogeneous Sedov-Taylor blast wave. Instead, the distribution resolved into two distinct kinematic populations:

\begin{itemize}

\item[(1)] A high-velocity group centered at $|$V$_\text{LSR}$$|$ $\sim$ 70\,$\pm$\,30 \kms. Given the advanced evolutionary stage of S147, these velocities are too high to represent the swept-up ISM unless the surrounding density is implausibly low. Therefore, we attribute this group to undecelerated ejecta within the low-density cavity, or ejecta that have been heated by reverse shock, which was launched inward when the forward shock impacted the dense cavity wall (see Section 4.2 for further discussion). 

\item[(2)] A low-velocity group peaking at $|$V$_\text{LSR}$$|$\,$<$\,20\,\kms. This population corresponds to the forward shock that has collided with and been stalled by the dense cavity wall, now propagating slowly as radiative shock.

\end{itemize}

Critically, this bimodality remains clearly evident even at small projected radial distances from the remnant's center (see Figure\,\ref{fig:fig_RV-position}, left panel), which rules out a purely geometric projection effect as the sole origin of the low-velocity component. The coexistence of these two kinematically distinct populations provides a direct observational constraint for the dynamical model of S147. In Section 4.2, we demonstrate that this bimodality is a natural prediction of the ``supernova-in-a-cavity'' scenario.

\subsubsection{Spatial Asymmetry and the ``Ear'' Morphology}

The spatial distribution of the blue-shifted (V$_\text{LSR}$\,$<$\,-20\,\kms) and red-shifted (V$_\text{LSR}$\,$>$\,20\,\kms) components (Figure\,\ref{fig:fig_spatial_distribution_snr-blue-middle-red}, left panels) exposes a marked radial asymmetry. The expansion is not isotropic; rather, it is dominated by protrusions aligned a preferential axis. Notably, the eastern ``ear'' structure, located at DEC between 27$^\circ$ and 28.5$^\circ$ and projected radial distances exceeding 100 arcmin from the geometric center, exhibits a strong concentration of blue-shifted emission, indicating that this portion of the remnant is expanding toward the observer at a higher velocity. 

Conversely, the western shell displays a more balanced mix of components. There is obvious asymmetry between the eastern and western ears. The western ear exhibits significantly fainter and more diffuse optical line emission compared to its compact and bright eastern counterpart. The observed optical asymmetry is consistent with the recent X-ray morphology reported by \citet{Khabibullin2024A&A...689A.278K}, where the X-ray emission reaches the boundaries of the optical and radio emission in the eastern part of the nebula, while ends slightly short on the western side.

\subsubsection{Expansion Velocity and Shell Geometry}

\begin{figure*}
\begin{center}
\includegraphics[scale=0.55]{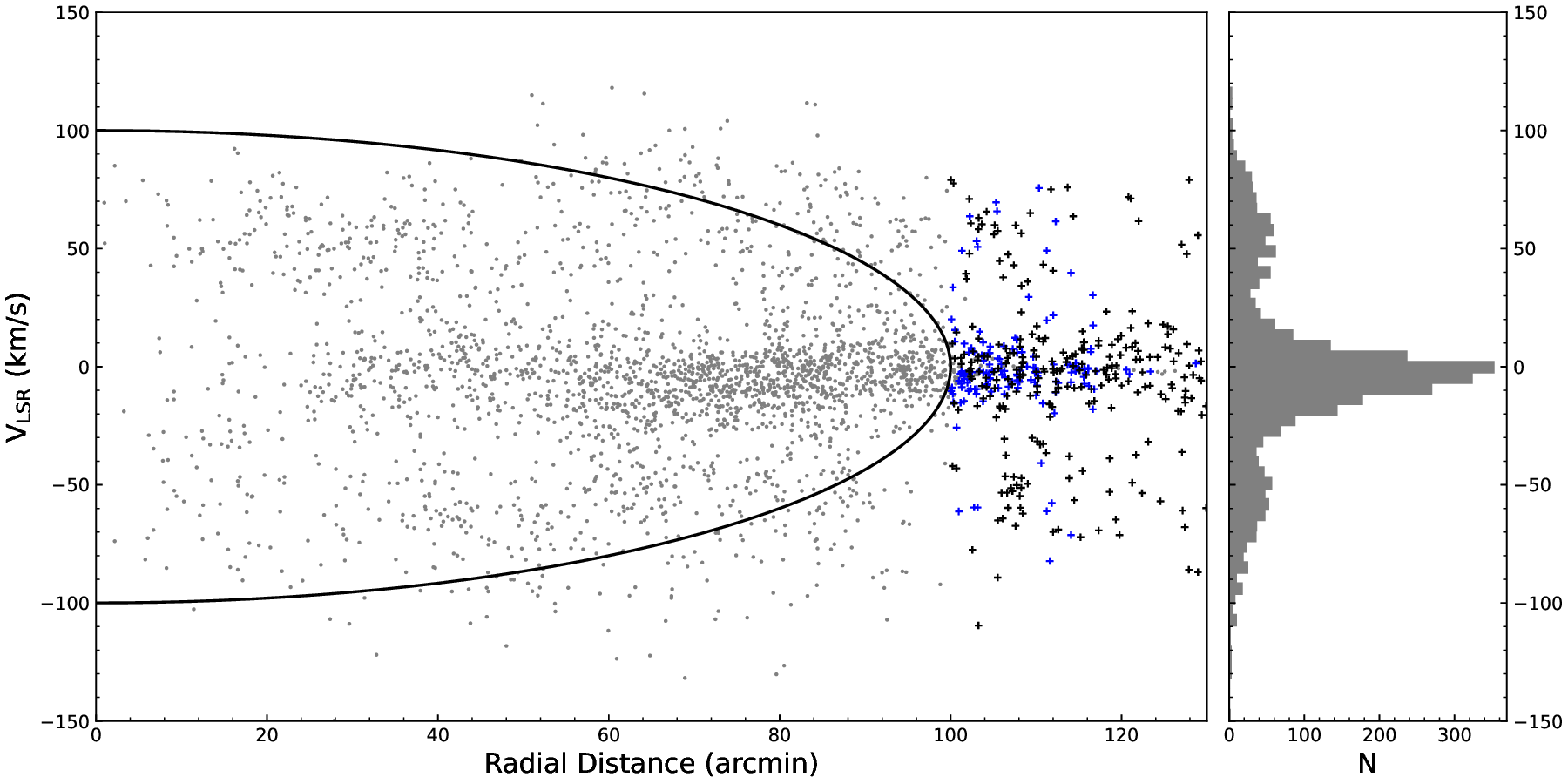} 
\caption{Left panel: Velocity-position diagram for the decomposed SNR components. The black solid line indicates a 100 \kms spherically expanding shell with 100 arcmin in radius. Black crosses denote the eastern ear region ($R$\,$>$\,100\,arcmin, 27$^\circ$ $<$\,DEC\,$<$\,28.5$^\circ$); blue crosses denote the northern boundary ($R$\,$>$\,100\,arcmin, DEC\,$>$\,28.5$^\circ$). The white circle in the first column of Figure\,\ref{fig:fig_spatial_distribution_snr-blue-middle-red} marks the 100 arcmin radius used to define the main shell extent. The right panel shows the corresponding histogram distribution of V$_\text{LSR}$.}
\label{fig:fig_RV-position}
\end{center}
\end{figure*}

Further insight into the global dynamics is provided by the velocity-position diagram (left panel of Figure\,\ref{fig:fig_RV-position}), which plots V$_\text{LSR}$ against the projected radial distance from the geometric center. The data points define an envelope consistent with a spherically expanding shell with a characteristic expansion velocity of $\sim$\,100\,\kms\ (the corresponding radius of $\sim$\,100\,arcmin is marked by the white circle in the first column of Figure\,\ref{fig:fig_spatial_distribution_snr-blue-middle-red}). Adopting the refined distance of 1.37 kpc, this implies a kinematic age of $t_{kin}$\,=\,$R/v$\,$\approx$\,0.39\,Myr assuming constant explosion (Section 4.2). Notably, this age is an order of magnitude larger than the kinematic age of the associated pulsar ($\sim$\,30\,kyr), suggesting that the observed $\sim$\,100\,\kms\ does not represent the average expansion velocity over the remnant's lifetime, but rather the current velocity of the shock front interacting with the cavity wall.

Furthermore, significant deviations from this smooth envelope are observed in the northern boundary (blue crosses in Figure\,\ref{fig:fig_RV-position}) and eastern ear (black crosses), where the shell appears to have broken out into a lower-density region of the ISM. This kinematic evidence confirms that S147 is transitioning into the momentum-conserving phase of its evolution, interacting with a highly inhomogeneous surrounding medium.

While the bulk of the SNR emission adheres to this $\sim$\,100\,\kms expansion envelope, we identify some points exceeding this limit, particularly within the radial range of 50\,--\,90\,arcmin (Figure\,\ref{fig:fig_RV-position}, left panel). These points predominantly correspond to discrete filamentary structures visible in Figure\,\ref{fig:fig_spatial_distribution_snr-blue-middle-red}. Physically, this deviation is physically interpreted as the forward shock interacting with an inhomogeneous cavity wall. The progenitor's stellar wind likely created a clumpy shell structure. When the forward shock encounters these high-density clumps, the local shock velocity can temporarily exceed the average expansion speed of the smooth component. Alternatively, these may represent the reverse shock propagating through the inner ejecta. Therefore, the $\sim$\,100\,\kms envelope remains a robust representation of the global dynamics, while the scatter reflects the complex microphysics of the shock-front interaction.

\section{Discussion}

\subsection{Shock Excitation Properties and the Eastern Jet}

The line-ratio diagnostics presented in Section 3.2 provide unambiguous evidence that the optical emission from S147 originated from radiative shock heating. The spatially resolved maps (Figure\,\ref{fig:fig_spatial_distribution_snr-dig} and \ref{fig:fig_spatial_distribution_snr-blue-middle-red}) reveal that the SNR filaments exhibit [\Ion{S}{II}]/\Ha\ ratios consistently exceeding 0.6, with values surpassing 1.3 in localized regions (such as, the eastern ``ear'' region). These ratios are characteristic of collisional excitation in fast shocks, where electron impact excitation dominates over recombination processes \citep{Fesen1985ApJ...292...29F}. In contrast, the separated DIG components display ratios typical of the warm ionized medium ($\sim$\,0.3\,--\,0.5), confirming the efficacy of our decomposition strategy.

A striking feature of the kinematic mapping is the pronounced asymmetry in the eastern sector of the remnant (black crosses in Figure\,\ref{fig:fig_RV-position}, Section 3.3.2). This region, corresponding to the ``ear'' structure at $R$\,$>$\,100 arcmin and DEC 27$^\circ$\,--\,28.5$^\circ$, exhibits the highest [\Ion{S}{II}]/\Ha\ ratios observed in S147, coupled with the blue-shifted radial velocities. This spatial coincidence of high excitation and high radial velocity strongly supports the presence of a directed outflow or jet. Following the classification scheme of \citet{Soker2023RAA....23k5017S}, S147 is a prime example of a ``one pair of ears'' core-collapse supernova remnant, where the morphology is shaped by long-lived jets launched during the final stages of the progenitor's evolution. The eastern ear represents the leading edge of a jet-driven outflow that has encountered a denser ambient medium, resulting in enhanced compression and shock excitation.

This interpretation is further corroborated by multi-wavelength observations. The X-ray morphology derived from SRG$/$eROSITA data \citep{Michailidis2024A&A...689A.277M} shows excellent spatial agreement with the optical filaments in this region, indicating that the shock front is sufficiently energetic to heat gas to X-ray emitting temperatures (see figure 2 in \citet{Michailidis2024A&A...689A.277M}). The kinematic and excitation properties of the eastern ear, therefore, provide direct evidence for the role of jet-induced shaping in core-collapse supernovae, linking the macroscopic remnant structure to the microphysics of the explosion mechanism.

However, a complete interpretation of the jet-driven scenario must account for the observed asymmetry between the eastern ear and its western counterpart. The pronounced asymmetry between the ears, with the western structure being fainter, more diffuse, and kinematically less coherent than its eastern counterpart, does not negate a common origin but rather provides critical insight into the ambient medium. This fragmented morphology of the western ear is corroborated by the recent finding that it exhibits ``two bright arcs on its boundary with an almost dark zone in between'' \citep{Shishkin2025ApJ...992..190S}. We interpret the pair as a bipolar outflow whose observable properties have been strongly modulated by an inhomogeneous circumstellar environment. The bright, well-defined eastern ear likely represents a jet component propagating into a lower-density region, preserving its kinetic energy to form a strong, confined shock. Conversely, the faint and diffuse western emission suggests its counterpart impacted a higher-density medium, leading to rapid deceleration, greater energy dissipation, and a subsequent faster cooling of the shock-heated gas. This interpretation of density-contrast-induced asymmetry is reinforced by the matching morphology in X-rays \citep{Khabibullin2024A&A...689A.278K}. Thus, the asymmetry itself serves as a diagnostic probe of the anisotropic mass-loss history of the progenitor system. While recent jet-based models attribute the ear morphology primarily to intrinsically asymmetric jets from the explosion itself \citep{Shishkin2025ApJ...992..190S}, our interpretation emphasizes the additional modulating role of a highly inhomogeneous circumstellar medium in shaping the observed brightness and kinematic contrast between the two ears.

\subsection{Dynamical Evolution: The "Supernova-in-the-Cavity" Scenario}

The kinematic structure revealed in Section 3.3, specifically the bimodal velocity distribution, presents a significant challenge to the classical Sedov-Taylor model of SNR evolution in a uniform medium. In a homogeneous environment, one would expect a unimodal distribution of radial velocities centered near the systemic velocity, reflecting a smoothly decelerating shell. Instead, our LAMOST MRS-N observations of S147 clearly resolve two distinct kinematic populations: a high-velocity group ($|$V$_\text{LSR}|$\,$\sim$\,70\,$\pm$\,30\,\kms) and a low-velocity group ($|$V$_\text{LSR}|$\,$<$\,20\,\kms).

This dichotomy is naturally explained by the ``supernova-in-the-cavity'' scenario, recently proposed to interpret the X-ray and optical properties of S147 \citep{Khabibullin2024A&A...689A.278K}. In this framework, the progenitor star, a massive O/B-type star, generated a powerful stellar wind during its main sequence lifetime. This wind swept up the surrounding ISM, excavating a low-density cavity surrounded by a dense, swept-up shell. The subsequent supernova explosion occurred within this pre-existing cavity.

The binary nature of the progenitor, implied by the surviving unbound companion HD~37424 \citep{Dincel2015MNRAS.448.3196D}, adds an important dimension to this scenario. Binary interaction is a common channel for massive star evolution, and the presence of a companion can substantially alter the progenitor's evolution and influence the structure of the surrounding circumstellar medium \citep{Petrovic2005A&A...435.1013P,Eldridge2008MNRAS.384.1109E, Ercolino2025A&A...696A.103E}. While \citet{Khabibullin2024A&A...689A.278K} demonstrated that a single-star wind-blown bubble can explain the X-ray properties of S147, the presence of a massive companion implies that the circumstellar environment may have been shaped by the combined action of two stellar winds. Such binary-wind interactions are known to produce asymmetric circumstellar structures in evolved stellar system \citep{Mastrodemos1999ApJ...523..357M,ElMellah2020A&A...637A..91E}, and by analogy, may have contributed to excavating a larger or more asymmetric cavity in the case of S147. However we caution that the degree to which HD 37424's wind contributed to excavating the observed cavity remains to be quantified with dedicated binary-wind simulations for massive progenitor system.

The observed velocity groups correspond to distinct physical components in this scenario: (1) The high-velocity group represents emission from undecelerated supernova ejecta or the reverse shock propagating back into the ejecta. Given the progenitor mass estimated to be 21.5\,--\,41.1\,$M_{\sun}$ \citep{Cruz-Cruz2026ApJ...999..162C}, the stellar wind was sufficiently powerful to excavate an extensive low-density cavity, allowing the ejecta to retain a significant fraction of their initial velocity. (2) The low-velocity group corresponds to the forward shock interacting with the dense shell of the cavity wall. Upon colliding with this dense interface, the shock slowed dramatically, producing the observed low-velocity, high-[\Ion{S}{II}]/\Ha\ emission characteristic of radiative shocks in dense media.

While the high-velocity component ($|$V$_\text{LSR}$$|$\,$\sim$\,70\,\kms) is a key prediction of the ``supernova-in-a-cavity'' scenario, our present data do not allow us to distinguish unambiguously between its two possible physical origins: reverse-shock-heated ejecta or freely expanding, cold ejecta that have not yet been overtaken by the reverse shock. Although kinematically similar, these two cases are expected to differ in their ionization state and spatial distribution. Future observations, particularly in the ultraviolet (to access high-ionization lines) or with higher spatial-resolution integral field spectroscopy, will be crucial to resolve this issue. Therefore, irrespective of its exact physical state, the detection of this high-velocity component provides important kinematic evidence for the propagation of ejecta through a low-density cavity, namely the wind-blown bubble created by the progenitor, thereby offering robust support for the presented physical picture.

This interaction radius is not arbitrary but is fundamentally set by the size of the pre-SN wind-blown bubble. Wind-blown bubble models have been extensively developed over decades. The canonical pressure-driven WBB model \citep{Castor1975ApJ...200L.107C,Weaver1977ApJ...218..377W} predicts that the shell radius evolves as $R$\,$\propto$\,$t^{3/5}$ during the energy-conserving phase. \citet{Chevalier1989ApJ...344..332C} further explored SNR evolution within pre-existing WBBs, showing that the remnant can be substantially decelerated upon encountering the dense swept-up shell.  Subsequent hydrodynamical simulations by \citet{Dwarkadas2005ApJ...630..892D,Dwarkadas2007ApJ...667..226D} demonstrated that core-collapse SNe expanding into circumstellar wind cavities undergo complex shock interactions upon encountering the dense swept-up shell, with the shock wave potentially trapped in the dense shell for several doubling times. These theoretical advances provide the foundation for interpreting the kinematics of S147.

For a $\sim$\,20\,$M_{\odot}$ progenitor with standard OB-star wind parameters, the canonical wind-blown bubble model, when accounting for a realistic total interstellar medium pressure of $\sim 10^{-12}$~dyn~cm$^{-2}$ (comprising thermal, magnetic, and cosmic-ray pressures), predicts a cavity radius of $R_{\mathrm{bubble}}$\,$\approx$\,38.5\,pc when a typical interstellar medium pressure is accounted for \citep[see the middle column of Table 2 in][]{Khabibullin2024A&A...689A.278K}. This theoretical scale, derived for the most physically motivated pressure environment, shows remarkable agreement with the observed physical radius of the \Ha\ shell in S147 ($\approx$\,40\,pc), providing independent, quantitative support for the scenario. The consistency between the predicted cavity size and the observed shock-front radius reinforces the interpretation that the current low-velocity, radiative shock is precisely located at the interface of the progenitor's cavity and its surrounding dense shell.

The global expansion velocity of $\sim$\,100\,\kms\ from the velocity-position diagram (Section 3.3.3) is consistent with the historical optical estimates \citep{Kirshner1979ApJ...229..147K,Phillips1981MNRAS.195..485P}. However, this velocity represents the current post-deceleration speed of the forward shock as it plows into the dense cavity wall, not its average speed over the remnant's lifetime.

This is underscored by a direct kinematic comparison. A naive assumption of constant expansion at the observed 100\,\kms, given the 40\,pc radius, would imply a kinematic age of $\sim$\,0.39\,Myr. This is in clear conflict with the robust $\sim$30\,--\,40\,kyr kinematic age of the pulsar derived from its 3D motion. Similarly, a naive application of the Sedov-Taylor self-similar solution for a point explosion in a uniform medium would yield an age of $\sim$\,150\,kyr \citep{Khabibullin2024A&A...689A.278K}. Moreover, adopting the true kinematic age of $\sim$\,30\,--\,40\,kyr and the observed radius of 40\,pc, the Sedov model predicts an expansion velocity of $\sim$\,400\,\kms\citep{Khabibullin2024A&A...689A.278K}, which is four times higher than the observed $\sim$\,100\,\kms. This order-of-magnitude discrepancy is naturally resolved within the ``supernova-in-a-cavity'' framework: the ejecta expanded freely at high speed through the low-density cavity, only decelerating to its current global expansion velocity of $\sim$\,100\,\kms\ upon collision with the dense shell. 

The observed bimodal velocity distribution provides direct kinematic evidence for this sequence of events. The high-velocity group ($|$V$_\text{LSR}$$|$\,$\sim$\,70\,\kms) is kinematically consistent with freely expanding ejecta or reverse-shocked material within the cavity interior. The low-velocity group ($|$V$_\text{LSR}$$|$\,$<$\,20\,\kms) corresponds to the forward shock that was decelerated abruptly upon colliding with the dense cavity wall. The consistency between the cavity model, the observed kinematics, and the independent pulsar age firmly establishes S147 as a middle-aged remnant in the crucial transition phase following its collision with the circumstellar shell.

\subsection{Impact on the Ambient Interstellar Medium}

The feedback from S147, as discussed in Section 4.1, is not limited to the direct interaction of the jets with the cavity walls. The remnant's global expansion also injects energy and momentum into the more diffuse surrounding ISM. Our analysis allows for a quantitative assessment of how the shock front interacts with and modifies the ambient DIG.

A key diagnostic of this interaction is the subtle but systematic elevation of the [\Ion{S}{II}]/\Ha\ ratio in peripheral DIG: while typical DIG in the Galactic plane exhibits [\Ion{S}{II}]/\Ha\,$\approx$\,0.3\,--\,0.5 \citep{Walterbos1992A&AS...92..625W}, the DIG surrounding S147's eastern and northern boundaries reaches $\sim$\,0.5\,--\,0.6 (Figure\,\ref{fig:fig_spatial_distribution_snr-dig}, lower panels). This enhancement cannot be attributed to photoionization from OB stars, which would produce lower [\Ion{S}{II}]/\Ha\ ratios. Instead, we detect a synchronous rise in [\Ion{N}{II}]/\Ha\  (0.5\,--\,0.6), confirming collisional excitation driven by shock compression \citep{Collins2001ApJ...551...57C}. The shock front compresses and heats the surrounding warm ionized medium as it breaks out of the cavity.

This interaction is particularly pronounced around the eastern ear (Section 4.1). The high-velocity jet-like outflow shaping this structure drives a lateral pressure wave into the surrounding medium, exactly the signature of the cavity-breakout process \citep{Khabibullin2024A&A...689A.278K}. The spatial correlation between the high-velocity SNR components and the elevated DIG line ratios provides direct evidence for feedback from the remnant to the ISM. The kinetic energy of the explosion is being thermalized and transferred to the surrounding gas, contributing to the turbulent support of the warm ionized medium.

Furthermore, the profound impact of S147 on its environment manifests across multiple wavelengths. While our LAMOST MRS-N observations directly trace the hydrodynamic feedback, specifically the compression and heating of the ambient DIG by the blast wave, recent high-energy observations reveal concurrent cosmic-ray feedback. Specifically, \citet{Yang2026ApJ..1009...10Y} detected hard-spectrum GeV $\gamma$-ray emission spatially correlated with adjacent molecular clouds (MCs), indicative of escaping cosmic rays interacting with dense target material. The fact that S147 is simultaneously driving a shock into the DIG (this work) and injecting accelerated particles into nearby MCs \citep{Yang2026ApJ..1009...10Y} underscores its multifaceted role in regulating the energetic and chemical budget of the ISM.

\section{Conclusions}

In this work, we presented the first dedicated medium-resolution spectroscopic study of the large-diameter SNR G180.0-01.7 (S147) using data from the LAMOST MRS-N survey. By performing multi-component Gaussian decomposition and applying combined spectroscopic and spatial criteria, we isolated the emission from the supernova remnant from the pervasive foreground and background DIG. Our primary findings are summarized as follows:

\begin{enumerate}
  \item We demonstrate that a combination of the [\Ion{S}{II}]/\Ha\ vs. [\Ion{N}{II}]/\Ha\ diagnostic diagram, kinematic filtering, and a morphologically defined spatial boundary provides a robust method for separating SNR shock-ionized gas from the pervasive DIG. This decomposition enables precise measurements of the remnant's kinematic and excitation properties.

  \item The radial velocity distribution of the SNR component reveals two distinct populations:  a low-velocity group ($|$V$_\text{LSR}$$|$\,$<$\,20\,\kms) from the forward shock stalled at the cavity wall, and a high-velocity group ($\sim$\,70\,$\pm$\,30\,\kms) from ejecta within the cavity. This co-existence is the characteristic kinematic signature of the ``supernova-in-a-cavity'' model, which reconciles the remnant's 40\,pc radius with the $\sim$\,30\,--\,40\,kyr kinematic age of its pulsar, pinpointing S147 in the transition phase immediately after cavity-wall collision.

  \item The eastern ``ear'' is unambiguously identified as a jet-driven outflow based on its systematic blue shift and locally extreme shock excitation ([\Ion{S}{II}]/\Ha\,$>$\,1.3). This confirms S147's classification as a ``one-pair-of-ear'' remnant shaped by an asymmetric explosion. Furthermore, the stark contrast between this bright, well-defined eastern structure and the faint, diffuse western ear illustrates how the observed asymmetry encodes information about the inhomogeneity of the pre-SN environment.

  \item We detect a systematic enhancement of the [\Ion{S}{II}]/\Ha\ ratio in the DIG along the remnant's northern and eastern peripheries. This finding provides direct evidence that the expanding shock front is actively compressing and collisionally exciting the adjacent warm ionized medium. This mechanical feedback complements recent high-energy feedback traced by gamma-ray observations of cosmic-ray interaction with nearby molecular clouds, establishing S147 as a coherent source of both mechanical and particle energy injection into its local Galactic environment.
\end{enumerate}

The questions raised here, including the exact physical state of the high-velocity ejecta and the detailed structure of the jet-cavity interaction, define clear targets for future high spatial-resolution integral field spectroscopy. The detection of the runaway companion HD~37424 suggests a binary origin for the progenitor, adding complexity to the single-star cavity scenario and raising the question of whether the cavity was shaped by binary-wind interactions. This possibility warrants further investigation through dedicated simulations. This work demonstrates that the medium-resolution, large-field spectroscopic surveys like LAMOST MRS-N are uniquely capable of decomposing the kinematics of large evolved SNRs and providing a powerful tool to link their morphology to the physics of their progenitors and explosions.

\section*{Acknowledgements}
We thank the anonymous referee for the helpful comments and suggestions that improved this manuscript. This work is supported by National Key R\&D Program of China grant No. 2025YFF0511000, the China Manned Space Program with grant no. CMS-CSST-2025-A19, National Natural Science Foundation of China (Nos. 12090041, 12090040), and the National Key R\&D Program of China grant (Nos. 2021YFA1600401 and 2021YFA1600400). This work is also supported by the National Astronomical Observatories of the Chinese Academy of Sciences, under Project No. E4TQ6801. 
This work made use of the data from LAMOST (Large Sky Area Multi-Object Fiber Spectroscopic Telescope, also known as the Guoshoujing Telescope) (https://cstr.cn/31118.02.LAMOST). LAMOST is a Chinese national mega-science facility, operated by National Astronomical Observatories, Chinese Academy of Sciences.

This paper makes use of data obtained as part of the IGAPS merger of the IPHAS and UVEX surveys (www.igapsimages.org) carried out at the Isaac Newton Telescope (INT). The INT is operated on the island of La Palma by the Isaac Newton Group in the Spanish Observatorio del Roque de los Muchachos of the Instituto de Astrofisica de Canarias. All IGAPS data were processed by the Cambridge Astronomical Survey Unit, at the Institute of Astronomy in Cambridge. The uniformly-calibrated band merged IGAPS catalogue was assembled using the high performance computing cluster via the Centre for Astrophysics Research, University of Hertfordshire.

\section*{Data Availability}

The data underlying this article are available from the corresponding author on reasonable request.

\label{lastpage}

\bibliographystyle{mnras}
\bibliography{ms} 

\appendix

\end{document}